Title:

# Unveiling dynamic changes in the diurnal microclimate of a *Buxus sempervirens* with non-intrusive imaging of flow field, leaf temperature, and plant microstructure

**Authors**:

Lento Manickathan [a, b, *]        (mlento@ethz.ch)

Thijs Defraeye [a, c]        (thijs.defraeye@empa.ch)

Stephan Carl [b]        (stephan.carl@empa.ch)

Henning Richter [d]        (henning.richter@uzh.ch)

Jonas Allegrini [b]        (ajonas@ethz.ch)

Dominique Derome [b]        (dominique.derome@empa.ch)

Jan Carmeliet [a]        (cajan@ethz.ch)

**Affiliations**:

[a] *Chair of Building Physics, Department of Mechanical and Process Engineering, ETH Zürich, Stefano Franscini Platz 1, CH-8093, Zürich, Switzerland*

[b] *Empa, Swiss Federal Laboratories for Materials Science and Technology, Laboratory of Multiscale Studies in Building Physics, Überlandstrasse 129, CH-8600, Dübendorf, Switzerland*

[c] *Empa, Swiss Federal Laboratories for Materials Science and Technology, Laboratory for Biomimetic Membranes and Textiles, Lerchenfeldstrasse 5, 9014, St. Gallen, Switzerland*

[d] *Diagnostic Imaging Research Unit (DIRU), Clinic for Diagnostic Imaging, Vetsuisse Faculty, University of Zurich, CH-8057, Zurich, Switzerland*

* corresponding author:

    Email:    mlento@ethz.ch

    Tel:    +41 (0)58 765 4604

    Address: Überlandstrasse 129, 8600 Dübendorf, Switzerland

**Highlight:**

- Leaf area density determined using X-ray tomography.
- Aerodynamic porosity does not reflect the true plant porosity distribution.
- Diurnal hysteresis between plant transpiration and leaf temperature is observed.
- Providing high-resolution dataset for numerical model validation.

# ABSTRACT

Plants modify the climate and provide natural cooling through transpiration. However, plant response is not only dependent on the atmospheric evaporative demand due to the combined effects of wind speed, air temperature, humidity, and solar radiation, but is also dependent on the water transport within the plant leaf-xylem-root system. These interactions result in a dynamic response of the plant where transpiration hysteresis can influence the cooling provided by the plant. Therefore, a detailed understanding of such dynamics is key to the development of appropriate mitigation strategies and numerical models. In this study, we unveil the diurnal dynamics of the microclimate of a *Buxus sempervirens* plant using multiple high-resolution non-intrusive imaging techniques. The wake flow field is measured using stereoscopic particle image velocimetry, the spatiotemporal leaf temperature history is obtained using infrared thermography, and additionally, the plant porosity is obtained using X-ray tomography. We find that the wake velocity statistics is not directly linked with the distribution of the porosity but depends mainly on the geometry of the plant foliage which generates the shear flow. The interaction between the shear regions and the upstream boundary layer profile is seen to have a dominant effect on the wake turbulent kinetic energy distribution. Furthermore, the leaf area density distribution has a direct impact on the short-wave radiative heat flux absorption inside the foliage where 50% of the radiation is absorbed in the top 20% of the foliage. This localized radiation absorption results in a high local leaf and air temperature. Furthermore, a comparison of the diurnal variation of leaf temperature and the net plant transpiration rate enabled us to quantify the diurnal hysteresis resulting from the stomatal response lag. The day of this plant is seen to comprise of four stages of climatic conditions: *no-cooling*, *high-cooling*, *equilibrium*, and *decaying-cooling* stages.

# 1. Introduction

The influence of plants on the microclimate in an urban environment is of growing interest due to the need of mitigating detrimental effects of urbanization and climate change on urban air temperature (Demuzere et al., 2014; Dimoudi and Nikolopoulou, 2003; Matthews et al., 2017; Shashua-Bar et al., 2009; Shashua-Bar and Hoffman, 2000; Yu and Hien, 2006). Plants modify the climate by intercepting solar radiation and by extracting heat from the environment through transpiration during photosynthesis (Nobel, 2009). Furthermore, the plant interferes with the airflow, extracting momentum and enhancing turbulent mixing (Finnigan et al., 2009; Gromke and Blocken, 2015; Sanz, 2003). Due to the present growing need to ensure that cities are resilient and can mitigate the rising temperatures, proposed mitigation strategies are to be properly assessed, and such assessment requires an adequate characterization of the effects of vegetation.

Foliage density is known to have an impact on the wind sheltering provided by a plant (Bitog et al., 2011, 2012; Guan et al., 2003). The aerodynamic properties of vegetation can be expressed simply through porosity and drag coefficient (Grant and Nickling, 1998; Guan et al., 2003; Manickathan et al., 2018a). The porosity and drag coefficient are known to depend on plant species (Cao et al., 2012; Manickathan et al., 2018a; Rudnicki et al., 2004; Vollsinger et al., 2005), age (Dahle and Grabosky, 2010) and, for deciduous plants that shed leaves during winter, the parameters have been observed to vary seasonally as well (Dellwik et al., 2019; Hwang et al., 2011; Maass et al., 1995). It is known that plant porosity can have an impact on turbulent mixing (Bai et al., 2012; Hiraoka and Ohashi, 2008; Manickathan et al., 2018a; McClure et al., 2017), which is seen to directly impact the thermal and pollutant dispersion characteristics of air flow (Conan et al., 2015; Gromke et al., 2008; Gromke and Blocken, 2015; Gromke and Ruck, 2007). Plants with high foliage density are seen to have a detrimental effect on pollutant dispersion of below-canopy pollutant sources such as automobiles (Nowak et al., 2006). Nevertheless, a high foliage density is also shown to have a beneficial impact on the pedestrian thermal comfort due to increased shading provided by the plant (Hwang et al., 2011; Morakinyo et al., 2017; Ng et al., 2012). Foliage density is parameterized using the leaf area index (LAI) to describe the net area of leaves and leaf area density (LAD) to describe the foliage distribution within the plant volume. These parameters are typically measured using optical techniques (Cao et al., 2012; Grant and Nickling, 1998; Guan et al., 2003; Liu et al., 2018; Manickathan et al., 2018a; Phattaralerphong and Sinoquet, 2005) that may compromise on the spatial accuracy and destructive techniques such as defoliation of the plant (Jonckheere et al., 2004; O'Neal et al., 2002). Solar radiation absorption within the foliage is known to depend primarily on the distribution of the leaf area density (Kichah et al., 2012; Manickathan et al., 2018b; Park et al., 2018). A dense plant canopy can result in a significant amount of solar radiation being absorbed resulting in a high leaf-to-air temperature (Hiraoka, 2005; Leuzinger and Körner, 2007; Manickathan et al., 2018b). Higher plant transpiration is then required to compensate for the high solar radiation absorption (Manickathan et al., 2018b), and this can result in a lower air temperature under the foliage (Wong et al., 2003). Studies have also revealed that plant transpiration rate varies not just due to atmospheric evaporative demand (AED) (Kichah et al., 2012; Manickathan et al., 2018b; McVicar et al., 2012; Tuzet et al., 2003) but can also dynamically vary during the day with higher transpiration during morning than in the evening (Huang et al., 2017; Tuzet et al., 2003). Therefore, experimental observations are key to determining the diurnal variability in the transpirative cooling performance of vegetation.

Various experimental approaches have been employed to assess the response of plants to environmental conditions ranging from field measurements (Dellwik et al., 2019; Grant and Nickling, 1998; Hagishima et al., 2007; Koizumi et al., 2016; Shashua-Bar et al., 2009; Shashua-Bar and Hoffman, 2000; Yuan et al., 2017), greenhouse studies (Fatnassi et al., 2006; Ganguly and Ghosh, 2009; Majdoubi et al., 2009; Montero et al., 2001), and wind tunnel experiments (Grace and Russell, 1977; Liu et al., 2018; Manickathan et al., 2018a; Miri et al., 2019; Rudnicki et al., 2004; Vollsinger et al., 2005; Yue et al., 2008). Wind tunnel experiments, which provide the most control over the airflow conditions, typically focus on the aerodynamic characteristics such as drag coefficient, porosity, and sheltering effect of plants and neglect the hygrothermal responses of the plant (Grace and Russell, 1977; Manickathan et al., 2018a). Therefore, the relationship between porosity heterogeneity and hygrothermal conditions of plants has yet not been to be experimentally observed and characterized. Moreover, few studies provide a high-resolution temporal and spatial study of the hygrothermal conditions of the plant which can be used for validating numerical models. Thus, there is a need for high-resolution experimental datasets investigating the links between plant morphology, and aerodynamic conditions including diurnal variations of the hygrothermal conditions such as air temperature, and relative humidity and how all these affect plant conditions.

The goal of the study is to experimentally quantify the influence of plant foliage geometry and environmental conditions such as wind speed and solar radiation on the transpirative cooling performance of a plant (*Buxus sempervirens*) inside a wind tunnel in a holistic approach. This is achieved by using multiple non-intrusive imaging techniques, thus measuring the plant foliage density with X-ray tomography, the wake flow field using stereoscopic particle image velocimetry (SPIV), the plant leaf temperature with infrared thermography and the hygrothermal conditions inside the foliage using various humidity and temperature sensors. The advantage of X-ray tomography to determine the plant porosity is that it is a non-intrusive approach to determining the plant structure. Thus the plant can undergo a series of additional experiments. This approach is inspired from the field of building physics where, for example, the determination of the microstructural morphology of building materials such as asphalt (Lal, 2016; Lal et al., 2017) or materials such as cotton textiles (Parada et al., 2017) is used to link the material configuration to its wetting and drying behavior. Thus, the approach allows us to quantify the impact of the plant foliage morphology on the wake flow characteristics, the hygrothermal conditions such as air temperature and relative humidity inside the plant foliage, the solar radiation penetration through the foliage and, finally, on the spatial distribution of the leaf temperature.

The study aims to answer the questions of how plant cooling varies spatially and temporally under variations of environmental conditions such as wind speed and solar radiation and whether a typical diurnal response of the plant could be defined. Moreover, the experiment provides a high-resolution dataset for future modeling validation studies. Given that this investigation is performed in a wind tunnel, we study the diurnal microclimate of a small plant. A large mature tree will most probably affect its environment in a dissimilar way. For example, the size of the plant, the flexibility of the branches and foliage, can influence the plant aerodynamic responses (de Langre, 2008; Manickathan et al., 2018a).

# 2. Methods and Materials

## 2.1 Material

The measurement campaign was performed for a small Buxus plant (*Buxus sempervirens*) in a wind tunnel as shown in **Fig. 1**. The plant foliage has a dimension $20 \times 20 \times 21$ cm$^3$ ($x \times y \times z$, i.e., streamwise, spanwise and vertical) as shown in **Fig. 1**b. The plant was placed in a pot, sealed using putty sealant to ensure water was lost only through leaf transpiration and no moisture is lost from the soil surface (**Fig. 1**c). The water loss due to transpiration was periodically compensated by adding fertilized water (1% (vol.) NPK 7-4-6 Buxus fertilizer). Prior to the experiment, the plant was periodically irrigated and exposed for one week to artificial sunlight with a 12-hour day-night cycle, where the day is a fixed-intensity photoperiod. The artificial sunlight was provided using an Osram Ultra-Vitalux 300 W solar simulator bulb, generating 13.6 W of UVA and 3.0 W of UVB. The bulb was placed 60 cm above the plant to provide 100 W m$^{-2}$ plant-canopy incident short-wave radiation. Furthermore, the growth of the plant foliage was periodically maintained to maintain the desired plant geometry as shown in **Fig. 1**. The detailed measurement of the plant morphology including geometry, porosity distribution, leaf size distribution, and total leaf area, was obtained using a high-resolution X-ray tomography measurement, as explained below.

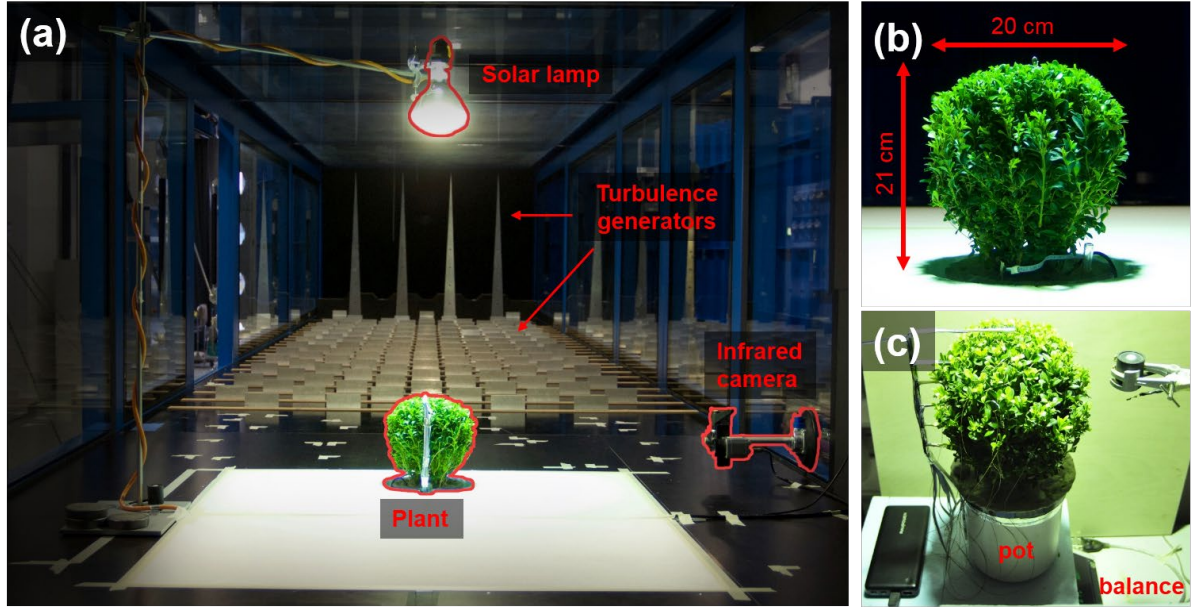

**Fig. 1** Wind tunnel setting for combined microclimate, SPIV, and infrared thermography measurements: a) Photo of plant installed in the tunnel, b) A close-up frontal view (windward) of the plant with dimensions, c) Photo of pre-experiment plant acclimation setup.

## 2.2 Experimental setup and procedure

The experimental campaign was divided into three stages: "*offline*" plant morphology measurement, pre-experimental controlled acclimatization setup, and the wind tunnel experiment. The "*offline*" measurement aimed to determine the plant morphological properties such as leaf area, porosity and porosity distribution using high-resolution X-ray imaging (**section 2.2.1**). The pre-experiment control setup aimed to acclimatize the plant to the wind tunnel boundary condition of solar radiation diurnal cycle. Furthermore, the sensors required for measuring the air and leaf temperature and air relative humidity within the foliage are mounted at this stage. Finally, the wind tunnel experiment aims at documenting the environmental conditions of the plant subjected to moderate wind. Thus, the flow field downstream of the plant, the hygrothermal microclimate inside the plant, the net transpiration rate and the plant foliage thermal profiles are measured.

## 2.2.1 X-ray imaging

A high-resolution computed tomography (X-ray CT) of the plant is acquired to determine the foliage morphology, the foliage porosity distribution, and the net leaf area. The advantage of such an approach is that it is a non-intrusive approach where the same plant can be further investigated (Lal et al., 2017; Patera et al., 2018). The measurement is performed at the Diagnostic Imaging Research Unit (DIRU) at the Vetsuisse Faculty, University of Zurich, using a Philips Brilliance CT 16-slice scanner, shown in **Fig. 2**, designed for medical imaging with an acquisition period of 39 seconds. The CT slices have a resolution of $0.318 \times 0.318$ mm$^2$ pixel with a slice thickness of 0.4 mm. The 12-bit image intensity data and the associated data of the measurement are stored in the DICOM file format. The resulting X-ray radiation intensity $I$ decay along the path $r$ is dependent on the initial intensity $I_0$ and the spatial distribution of the attenuation coefficient $\mu$ along the path $r$, given by the Beer-Lamberts law:

$$I(r) = I_0 \exp\left\{-\int_0^r \mu(r)\,\mathrm{d}r\right\} \quad (1)$$

The plant properties such as net leaf area are obtained from the 3D dataset after image processing, consisting of image enhancement, image segmentation, and classification.

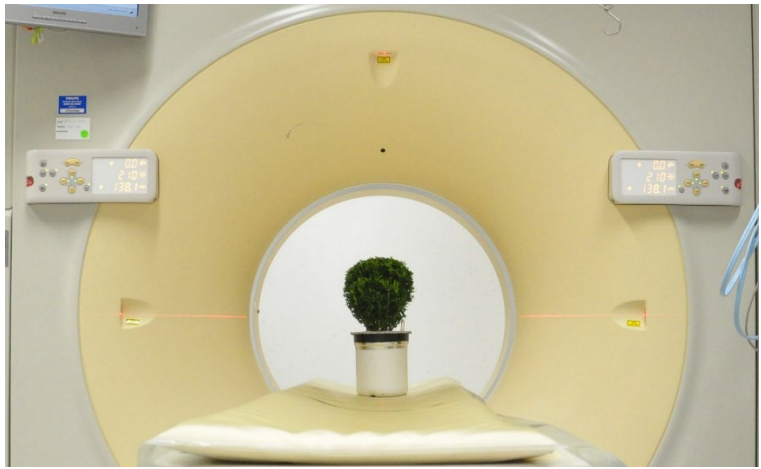

**Fig. 2** X-ray imaging setup of the plant specimen at Diagnostic Imaging Research Unit (DIRU) at the UZH. The specimen is imaged with Philips Brilliance CT 16-slice scanner, a medical imaging device.

## X-ray computed tomography

The attenuation coefficient indicates the absorption of the biological material to X-ray radiation, indicating the variability in the biological composition of the sample. The reconstructed tomographic data provides the 3D distribution of X-ray attenuation by the sample. The Hounsfield scaling normalizes the X-ray attenuation coefficient $\mu$ with that of the air $\mu_{air}$ and distilled water $\mu_{water}$ at standard atmospheric conditions, where the Hounsfield units of air and water are $HU_{air} = -1000$ and $HU_{water} = 0$, respectively:

$$HU = 1000 \times \frac{\mu - \mu_{water}}{\mu_{water} - \mu_{air}} \quad (2)$$

HU can be used to extract the biological properties of the plant as the scaling can be a means for fast and simple categorization of biological matter with different water quantity.

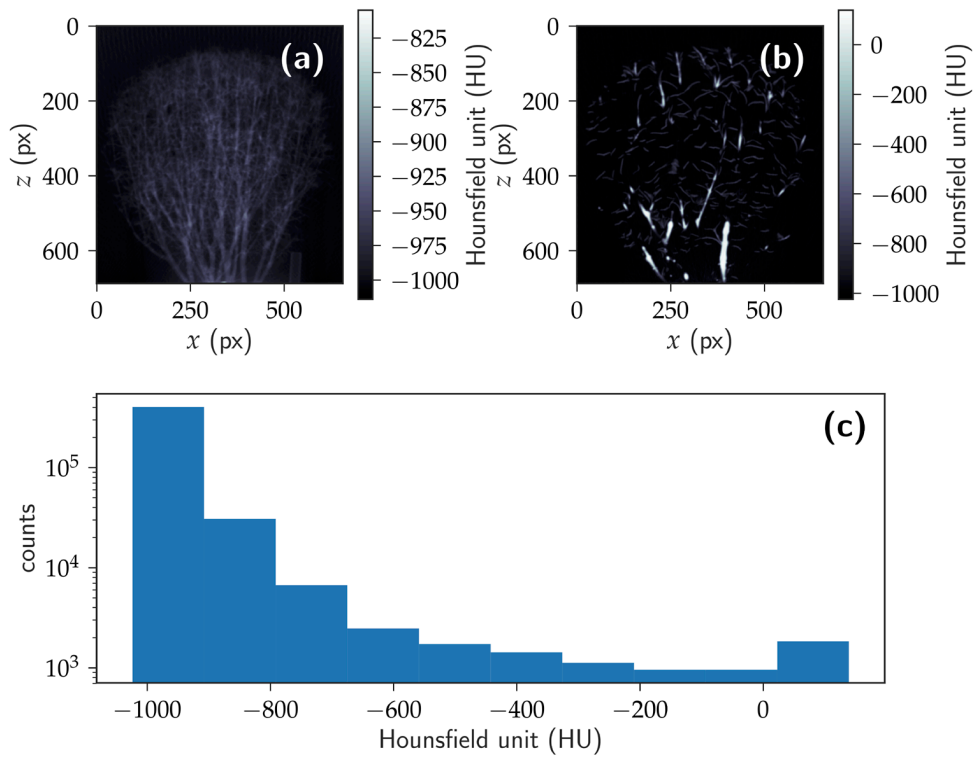

**Fig. 3** Raw image obtained from X-ray CT-scan: a) the side view of CT-Scan, b) a single image slice at the middle of the plant, c) Histogram of (b). The plant attenuation is represented in Hounsfield units (HU) where $-1000$ corresponds to air and 0 corresponds to pure water. The image slice resolution is $\Delta x = \Delta z = 0.318$ mm with a slice thickness of $\Delta y = 0.4$ mm.

*Image segmentation*

**Fig. 3**a shows a side view of the CT-scan, whereas **Fig. 3**b shows a single image slice at the middle of the plant. **Fig. 3**c shows the histogram distribution of the single image slice. A preliminary observation of the images is that air, leaf, and branch show distinct attenuation coefficients for which manual thresholding of the histogram is performed. Based on the histogram, a simple segmentation assigns air, leaves, and branches. The pixels are assigned as for air ($-1000$ to $-900$), leaves ($-900$ to $-700$), branches ($-700$ to 200), and anything beyond as foreign material. The bounds are determined with the help of $k$-means clustering method assuming a tri-modal distribution (i.e., $k = 3$) of the histogram. The histogram of X-ray CT scan (**Fig. 3**c) shows that this assumption is not valid, however, this histography-based segmentation serves as a base case for the more advanced segmentation algorithm. **Fig. 4**a shows the original dataset before the classification and **Fig. 4**b shows the histogram-based classification labels. A first observation shows a reasonable classification of the dataset except at boundaries of the branches, due to lower attenuation at the boundaries of the objects.

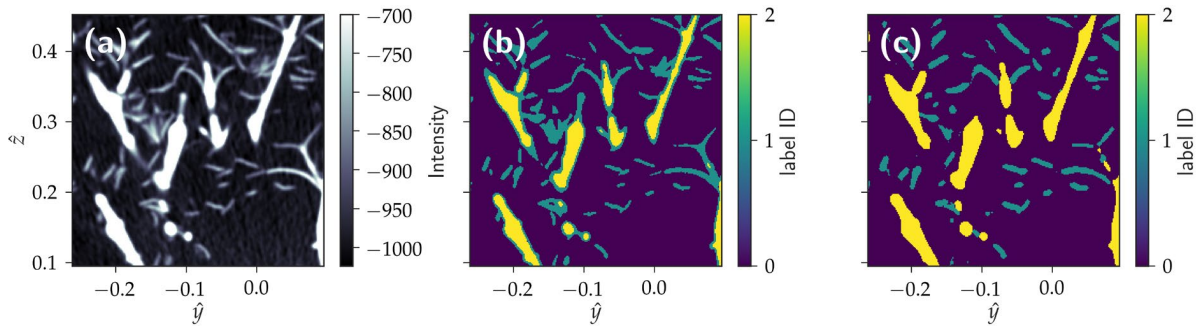

**Fig. 4** Segmentation of the X-ray CT scan: a) slice of original X-ray CT dataset, b) segmentation using user-defined histogram threshold and c) segmentation using Trainable WEKA Segmentation and additional morphological operation (opening + closing). Only a sub-region of an image slice is shown for clarity. The segmented pixels are labeled as air (0, purple), leaf (1, blue) and branch (2, yellow).

Therefore, a more advanced approach, the Trainable Weka Segmentation (TWS) (Arganda-Carreras et al., 2017) and additional binary morphological operations are used for classification of the three components of the tree. The TWS segmentation employs an implementation of fast random forest ensemble classification algorithm in the Fiji application (Schindelin et al., 2012) using 200 decision trees with two random features from user selection. A larger selection of edge-enhancement filters is used to reduce the over-estimation of the leaves at the boundary. The training procedure consisted of providing an initially labeled training dataset, trained using the classifier, visually validating the classification, and improving the user-provided labeled training dataset for improved classification. Finally, small remaining leaf pixels at the boundary of the branches were removed using additional morphological operations ("opening" + "closing" (Haralick et al., 1987)) using the python library, *scikit-image* (van der Walt et al., 2014). **Fig. 4**c shows the resulting segmentation using the decision tree classification and morphological operation. The figure shows a better classification of the plant elements. From this processed dataset, the plant surface mesh for branches and leaves can be generated to obtain metrics such as net leaf surface area.

### 2.2.2 Wind tunnel setup

The measurement setup is a holistic measurement approach for simultaneously measuring multiple microclimate parameters. The climatic variable measurements consisted of the measurement of the air relative humidity, air temperature, the net plant transpiration, and the flow velocity. The measurements were performed inside the ETHZ / Empa Atmospheric Boundary Layer (ABL) wind tunnel, a closed circuit Göttingen type wind tunnel with a test section cross-section of $1.9 \times 1.3$ m$^2$ ($W \times H$). The blockage ratio (i.e., the frontal area of the plant to wind tunnel cross-section) of the plant is determined to be 1.7% and is, therefore, neglected. **Fig. 1**a shows the wind tunnel setup that is employed to ensure minimal disturbances from measurement instruments. During the microclimate measurement, the diurnal variations of the air temperature and relative humidity at different heights were recorded using RH/T (i.e., combined relative humidity and temperature measurement) sensors. In addition, the leaf temperature was measured using infrared thermography. The net transpiration rate from the plant was measured using a mass balance positioned below the wind tunnel floor through an access panel, ensuring that the disturbance of the air flow is affected by the presence of the plant only (**Fig. 6**b). The airflow leeward of the plant was measured using stereoscopic particle image velocimetry (SPIV).

*Microclimate boundary conditions*

A parametric study on the steady-state and dynamic response of the plant exposed to four environmental conditions was performed: two wind tunnel set wind speeds, 0 and 1 m s$^{-1}$, and two plant-canopy incident solar radiation levels, 0 and 100 W m$^{-2}$. The air temperature and the relative humidity inside the wind tunnel were 21 °C and 25% RH, where the condition was coarsely regulated by the HVAC system of the wind tunnel facility.

**Fig. 5** shows the vertical profiles of the mean upstream streamwise velocity and the Reynolds stresses measured in the empty tunnel at the future position of the plant. At plant canopy height, $H = 210$ mm, the measured mean velocity is $U_H = 0.77$ m s$^{-1}$ for the wind speed of $U_{ref} = 1$ m s$^{-1}$. The wind tunnel flow was modified using turbulence generators, as shown in **Fig. 1**a, to generate an appropriate ABL profile typically found in an urban context (Tsalicoglou et al., 2018). **Fig. 5** shows the mean velocity and variance of velocity profiles that is typical for an ABL, with a turbulent intensity $I = \sqrt{2/3 \cdot k}/U_{ref} = 12.1\%$. This turbulence intensity thus exposed the plant foliage to air flow characteristics of an urban boundary profile.

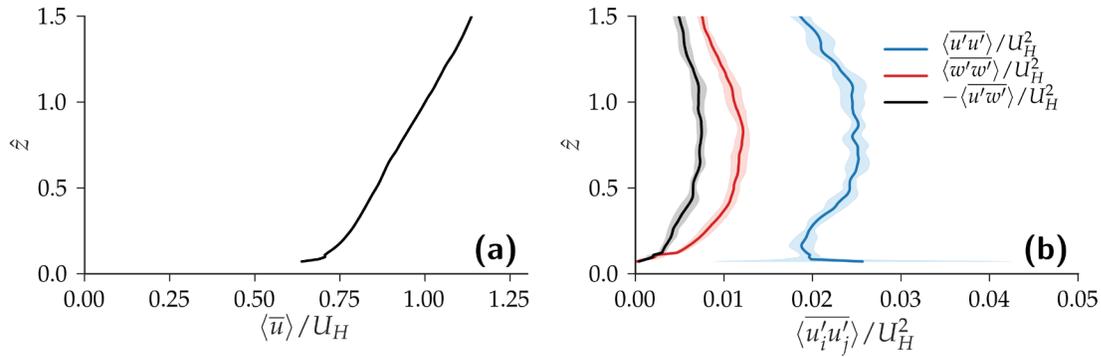

**Fig. 5** Vertical profiles of incoming a) normalized mean streamwise velocity $\langle \bar{u} \rangle / U_H$ and b) normalized Reynolds stresses $\langle \overline{u_i' u_j'} \rangle / U_H^2$. The normalized vertical height $\hat{z} = z/H$. The canopy velocity at $H = 210$ mm is $U_H = 0.77$ m s$^{-1}$ ((Tsalicoglou et al., 2018).

*SPIV setup*

A stereoscopic particle image velocimetry (SPIV) setup was used to measure the time-averaged 3D wake structure of the plant. To reconstruct the full 3D wake of the plant, the SPIV setup is traversed eight times vertically from 60 mm from the ground upwards to 270 mm, at 30 mm intervals, to produce 8 SPIV planes, as depicted in **Fig. 6**a. The time-averaged measurements of each plane are then combined to generate a 3D velocity field of the plant wake flow and its associated turbulence statistics.

The SPIV measurement setup consists of two 2560 × 2160 pixel s-CMOS HiSense Zyla camera and a 200 mJ/pulse (at 15 Hz) Nd-YAG Litron laser which is traversed together using a high-precision system. The two cameras are placed at 38° and 0° normal to the imaging plane. The wind tunnel is seeded with 1 $\mu$m Di-Ethyl-Hexyl-Sebacat (DEHS) tracer particles and the velocity vectors are calculated using an iterative cross-correlation algorithm of Dantec DynamicStudio with final interrogation area of $32 \times 32$ px$^2$ and 50% overlap. The field of view (FOV) of a SPIV plane is $438 \times 559$ mm$^2$ and provides 69 015 PIV vectors per plane with an in-plane resolution of 2.5 mm, whereas the plane-to-plane resolution is 30 mm. To obtain statistically relevant turbulence characteristics, 5000 images are obtained at 15 Hz. Furthermore, to ensure low optical inferences during the SPIV measurement, the hygrothermal measurement devices, thermal imaging devices, and the solar lamp had to be removed. The two distinct setups for airflow measurement and hygrothermal measurement are shown in **Fig. 6**a and **Fig. 6**b, respectively. Furthermore, we assume that the influence of removed devices is minimal for low wind speeds, such as in our case.

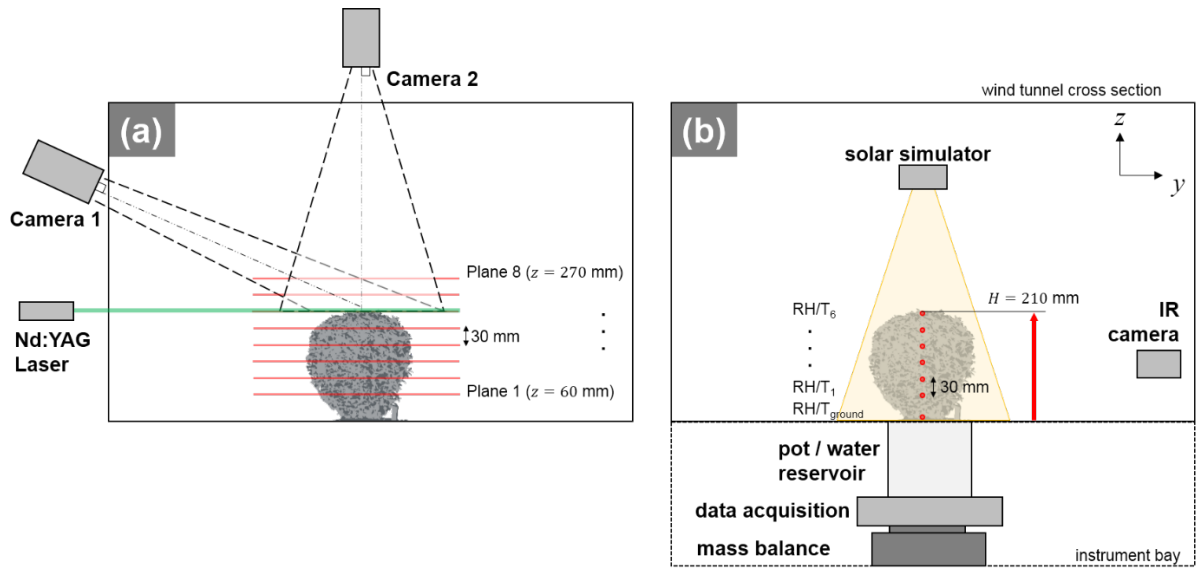

**Fig. 6** Schematic cross section representations of wind tunnel setups used to measure the airflow and hygrothermal microclimate of the plant: a) SPIV setup for multi-plane time-averaged measured velocity consisting of 8 horizontal planes, $z =$ [60, 90, 120, 150, 180, 210, 240, 270] mm, b) microclimate measurement setup consisting of IR camera and SHT sensors inside the wind tunnel, mass balance and other data acquisition system below the wind tunnel through an access panel. We note the presence of the solar simulator in the second setup. The two distinct setups are required to attain non-interfered optical measurements of the SPIV.

*Microclimate and net transpiration rate measurements*

The hygrothermal microclimate conditions within the foliage at various wind and radiation conditions are investigated separately from the flow field measurements, as clarified in **Fig. 6**. **Fig. 6**b shows the setup used to measure the hygrothermal microclimate of the plant. The solar simulator is placed above the plant providing 100 W m$^{-2}$ of incident short-wave radiation at the plant canopy ($H = 210$ mm). The solar simulator is controlled using a time switch that provides 12-hour periods of 0 and 100 W m$^{-2}$ radiation in alternance, imposing a simplified diurnal solar cycle. A fixed radiation intensity was imposed to obtain the steady-state response on the plant. In the future, a time-dependent intensity profile such as sine profile can be investigated. The net plant transpiration is measured using a Mettler PM6100 mass balance placed below the wind tunnel floor in the instrument bay along with the plant pot and water reservoir and the data acquisition system, to minimize their inferences on the flow field. The mass balance has a maximum capacity of 6.1 kg with an accuracy of ($\pm$ 0.1 g). Simultaneously, the vertical relative humidity and temperature profiles inside the plant are measured using Sensirion SHT35 sensors, as indicated in **Fig. 6**b. Seven SHT sensors are placed as follows: five sensors are directly inside the foliage with a vertical offset of 30 mm starting below the plant canopy (i.e., $z = [60, 90, 120, 150, 180]$ mm), one sensor is at plant canopy (i.e., $z = 210$ mm), and one sensor is below the plant foliage near the ground (i.e. $z = 0$ mm), to measure the shaded conditions. An eighth sensor is placed directly downstream of the plant at the height of the plant canopy (i.e., $z = 210$ mm). The eight SHT sensors have an accuracy of $\pm 1.5\%$ RH (between 0 and 80% RH) and $\pm$ 0.1 °C (between 20 and 60 °C). All the sensors for the plant are connected and powered by the wireless data acquisition system directly below the plant pot, such system is necessary to not hinder the mass balance measurement of water loss due to transpiration. The wireless data acquisition system consists of an Arduino Micro and a 20 100 mAh powerbank providing the necessary power for a multi-day measurement period. The Arduino Micro serves not only as an analog-to-digital signal converter but also as a telemetry device for sending the acquired data to the data logger away from the measurement setup. The data is acquired at a 30-second interval.

*Infrared imaging*

Infrared (IR) thermography is performed to obtain a high-resolution spatial and temporal data of the foliage temperature when exposed to varying environmental conditions. The feasibility of employing infrared thermography to obtain leaf temperature variability has been demonstrated in the past (Jones, 1999; Merlot et al., 2002). IR imaging system is employed to measure the outer plant foliage temperature simultaneously with the hygrothermal and net transpiration rate measured for two different wind conditions throughout the diurnal radiation cycle. The IR imaging system consists of the Optris PI 640 IR camera with a $640 \times 480$ px$^2$ sensor, a spectral response between 7.5 and 13 $\mu$m and is set to measure $-20$ to 100 °C with an accuracy of ± 2 °C (Allegrini, 2018; Tsalicoglou et al., 2018). A 33° lens is used providing an effective FOV of $223 \times 211$ mm$^2$. The IR measurement is performed at a frequency of 1 frame/minute throughout the diurnal cycle. A PT100 (platinum resistance thermometer) sensor is placed in FOV of the IR image for calibration (Allegrini, 2018).

# Results and Discussion

## 3.1 X-ray tomography: A non-destructive approach to obtain plant traits

### 3.1.1 Plant porosity distribution

**Fig. 7**a shows the average porosity for representative elementary volumes (REV) of voxel sizes ranging from 5 to 100 px$^3$. To calculate the spatial distribution of plant porosity, a sufficiently large REV has to be chosen to ensure that variability in the plant microstructure is taken into account. However, if the REV is too large, it sacrifices the resolution of the porosity distribution. If the REV size is 100 px$^3$, the resulting porosity is $\langle \phi \rangle_{xyz} = 90\%$. However, the downside of such a large REV size is that it has scarified the spatial resolution of porosity distribution. Therefore, in our case, an REV size of 30 px$^3$ was seen to be optimal, providing sufficient accuracy and resolution, with a calculated average plant porosity of $\langle \phi \rangle_{xyz} = 88\%$ and error of only 2%. **Fig. 7**a also shows the average porosity obtained from both "weka" and the "hist." methodology as described in section 2.2.1. We see that both approaches follow the same profile with minor differences. This result indicates that both approaches accurately differentiates air space and biological material. However, in section 2.2.1, it was seen that the "weka" method provides a better classification of leaf and branches, as visible in **Fig. 4**. So, to obtain key parameters such as leaf surface area, the "weka" method is opted.

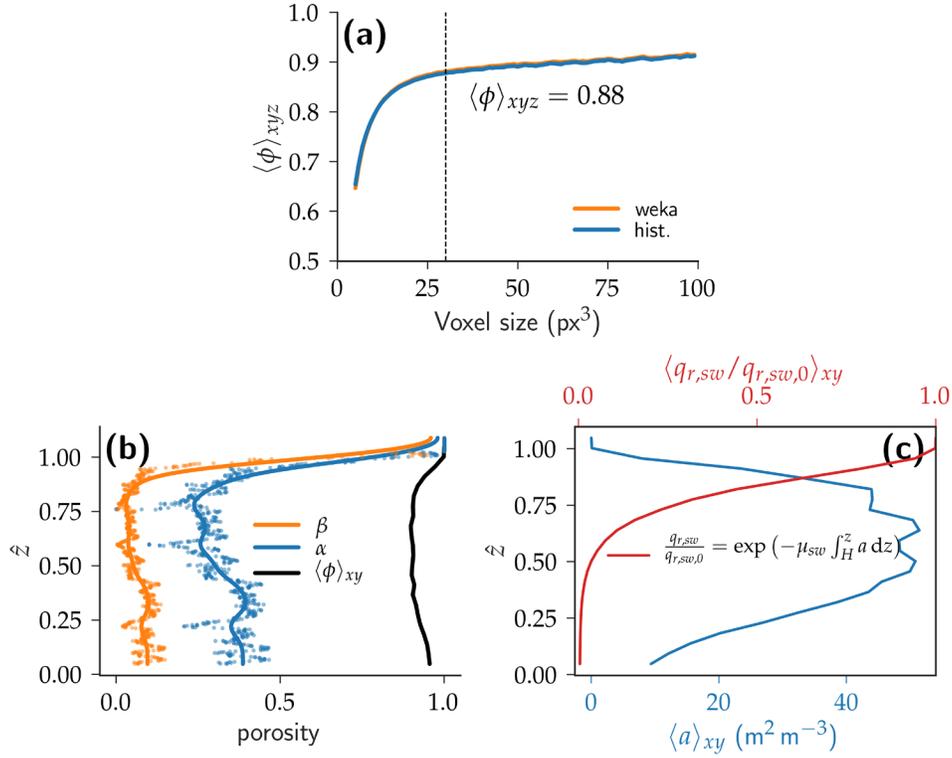

**Fig. 7** a) Average plant porosity $\langle\phi\rangle_{xyz}$ with respect to voxel size (px³) as a base for determining the REV for calculating porosity distribution, b) Three vertical porosity distributions: optical $\beta$, aerodynamic $\alpha$ and true porosity $\langle\phi\rangle_{xy}$, and c) leaf area density $\langle a\rangle_{xy}$ (m² m⁻³) (blue) and normalized short-wave radiative flux $q_{r,sw}$ profile (red) inside vegetation determined using Beer-Lambert law extinction coefficient of short-wave radiation $\mu = 0.78$ (Manickathan et al., 2018b).

**Fig. 7**b shows three vertical distributions of plant porosity. In conjunction with the plant porosity based on the X-ray CT (black line), the optical and aerodynamic porosity of the plant (with respect to the incident airflow direction in the wind tunnel) is investigated. The optical and aerodynamic porosity are typical measures used in the wind tunnel studies to estimate the aerodynamic contribution of the plant porosity (Grant and Nickling, 1998; Guan et al., 2003; Manickathan et al., 2018a). The optical porosity $\beta$ and the aerodynamic porosity $\alpha$ can be related as follows:

$$\alpha = \beta^{0.4} \tag{3}$$

empirically derived from wind tunnel measurements (Guan et al., 2003). The optical porosity is obtained from a 2D optical image of the tree perpendicular to the incident flow direction (**Fig. 8**a) and is defined as the ratio of empty pixels (without plant elements) to the total number of pixels within the silhouette of the plant, shown in **Fig. 8**. A convex hull is used to define the silhouette of the plant. Furthermore, the streamwise-averaged plant porosity $\langle\phi\rangle_x$ (**Fig. 8**b) is investigated to quantify how much the plant that is blocking the flow field at a given location. It indicates clearly that the highest blockage is found near the top half of the plant (i.e., $\hat{z} = 0.75$). Such location of high plant foliage density can lead to high solar radiative absorptions and a resulting high transpiration rate, a hypothesis that will be studied by measuring the air temperature and relative humidity, showing the regions of high transpiration. The plant foliage density is seen to reduce towards to the edges gradually and at these regions, a lower blockage might be present with a higher bleed flow, which can provide internal ventilation, increasing the convective dominated processes such as sensible and latent heat fluxes (Manickathan et al., 2018b).

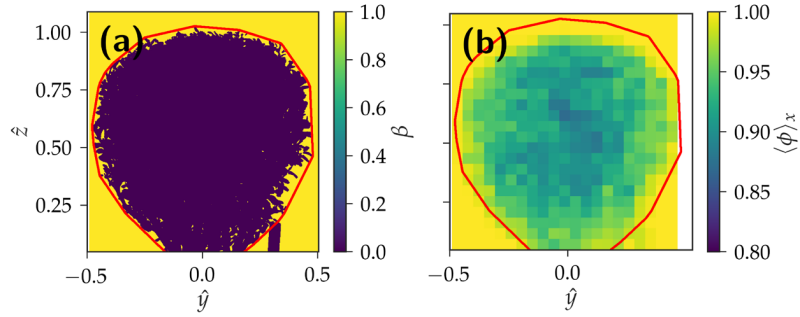

**Fig. 8** Porosity distributions of the plant: a) The optical porosity $\beta$ with plant element and airspace indexed as 0 and 1, respectively and b) streamwise-averaged plant porosity $\langle\phi\rangle_x$.

The horizontal-averaged ($x - y$ plane) porosity distribution is shown and compared in **Fig. 7**b. Comparing optical, aerodynamic and real 3D-based porosity (i.e., the plant porosity), we see that the real 3D-based porosity of the plant is substantially higher than the optical and the aerodynamic porosities. This indicates that the aerodynamic porosity usually used in various studies (Bitog et al., 2011; Guan et al., 2003; Manickathan et al., 2018a), does not reflect the true plant porosity. However, it is still uncertain as to which porosity provides the best description of the impact of the tree on the flow field such as the wake velocity deficit and the turbulent kinetic energy profile. Therefore, the influence of true plane porosity $\phi$ and the aerodynamic porosity $\alpha$ of the wake velocity statistics is investigated in **section 3.2**.

### 3.1.2 Plant surface mesh and total leaf area

In addition to the porosity distribution of the plant, the total leaf area, leaf area index and leaf area density are important parameters for understanding and modeling the influence of vegetation (Manickathan et al., 2018b). The plant parameters are obtained from X-ray tomography, and the surface of the plant components is generated from the volumetric data of the plant X-ray CT. **Fig. 9**a shows the leaf and branch surface colored as green and orange, respectively. **Fig. 9**b shows an internal sub-volume of the plant for clarity. The surface geometry is generated using a marching-cube algorithm implemented in *scikit-image* (van der Walt et al., 2014).

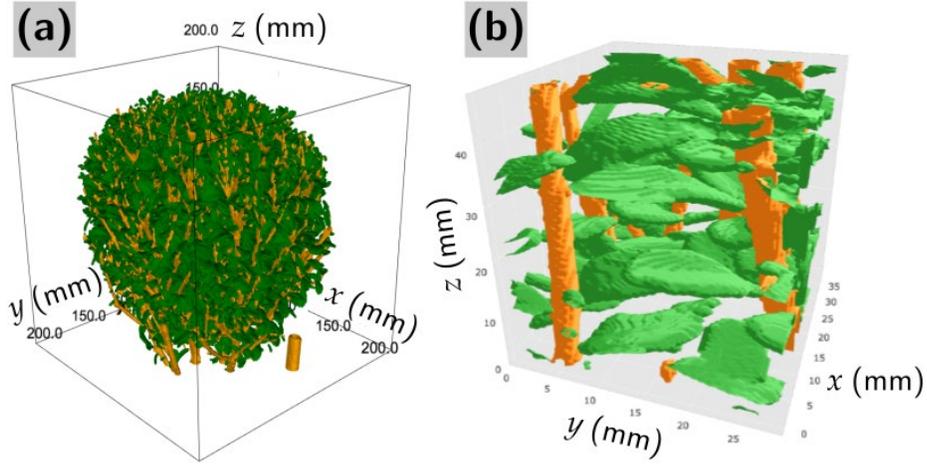

**Fig. 9** Surface geometry of the plant, where leaves (green) and branches (orange) are differentiated: a) complete plant surface, b) sub-volume inside the plant foliage.

The total leaf surface area is determined by integrating the mesh surface and is calculated to be $A_l = 0.75$ m². A metric of plant trait commonly used to quantify the amount of leaves is the leaf area index (LAI). The leaf area is defined as the ratio of one-sided leaf area to the plant ground cover area $A_g$. The one-sided leaf area is simply assumed to be half the total measured leaf area, and the plant ground cover area is derived from the X-ray CT dataset. The plant ground cover was determined to be $A_g = 0.031$ m², and a resulting leaf area index of $LAI = 12.14$ m² m$^{-2}$ is measured. Therefore, the plant is seen to have a high LAI because it is a shrub species that is commonly used as hedgerows or shelterbelts.

The (one-sided) leaf area density $a$ (m² m$^{-3}$) (or also known by LAD) is a key parameter that is used to represent vegetation as a porous medium in a numerical models (de Langre, 2008; Gross, 1987; Manickathan et al., 2018b). It is related to the net leaf surface area (two-sided) $A_l$ and the plant porosity $\phi$, given as:

$$a = \frac{1}{2} A_l \frac{1-\phi}{\int 1-\phi \, dV} \quad (4)$$

**Fig. 7**c shows the vertical distribution of the horizontally-averaged (i.e., streamwise $x$ and spanwise -averaged $y$) leaf area density $\langle a \rangle_{xy}$ (m² m⁻³). The $a$ has a peak value of $\langle a \rangle_{xy} \approx 50$ m² m⁻³ where the porosity is at lowest and approaches zero at the top and bottom regions of the foliage as $\phi \to 1$. An important aspect of the $a$ distribution is its influence on the solar radiation attenuation within the foliage. The extinction of solar radiation within the foliage due is typically modeled using a simple Beer-Lambert law and can be assumed to depend only on the $a$ distribution and the extinction coefficient of the short-wave radiation $\mu_{sw}$ (Manickathan et al., 2018b). **Fig. 7**c also shows the vertical distribution of normalized short-wave radiative flux $\langle q_{r,sw} \rangle_{xy}$ determined from Beer-Lambert law. We see that nearly 50% of the above-canopy solar radiation intensity $q_{r,sw,0}$ is absorbed within just the top 20% of the plant foliage and a further 30% within the next 10% of the foliage. Therefore, most of the solar radiation is seen to be absorbed within the top layer of the plant indicating that the influence of solar radiation will be present at this region (Manickathan et al., 2018b). To investigate this further, the influence of measured leaf area density distribution on the hygrothermal climate variables ($T, T_l, RH$) inside the foliage is investigated in **section 3.3**.

## 3.2 3D wake flow characterization by SPIV

The 3D wake flow field is studied using stereoscopic particle image velocimetry (SPIV). The study aims to understand the influence of the porous plant microstructure on the wake flow characteristics. Therefore, the mean velocity and the turbulence kinetic energy (TKE) of the plant wake is studied and compared with the porosity distribution. The setup of the PIV system is detailed in **section 2.2.2**, measuring eight horizontal planes behind the plant at vertical heights of z = [60, 90, 120, 150, 180, 210, 240, 270] mm, heights normalized below to the height of the plant.

### 3.2.1 Normalized mean velocity magnitude

The time-averaged mean velocity of the plant wake is studied first to understand how the plant porosity influences the flow. **Fig. 10** shows the normalized mean velocity magnitude $|\bar{u}|/U_H$ where $U_H = 0.77$ m s⁻¹. The coordinate system is normalized with the tree height of $H = 210$ mm, where $\hat{x}_i = x_i/H$ for $x_i = \{x, y, z\}$. This mean velocity distribution shows a prevailing recirculating flow for $\hat{z} \leq 0.71$, whereas above $\hat{z} > 0.71$, the streamlines indicate a bleed flow with $\phi \to 1$ (**Fig. 8**b). To further link the flow field to the climate, the influence of the flow conditions on the hygrothermal microclimate inside the plant will be investigated in **section 3.3**.

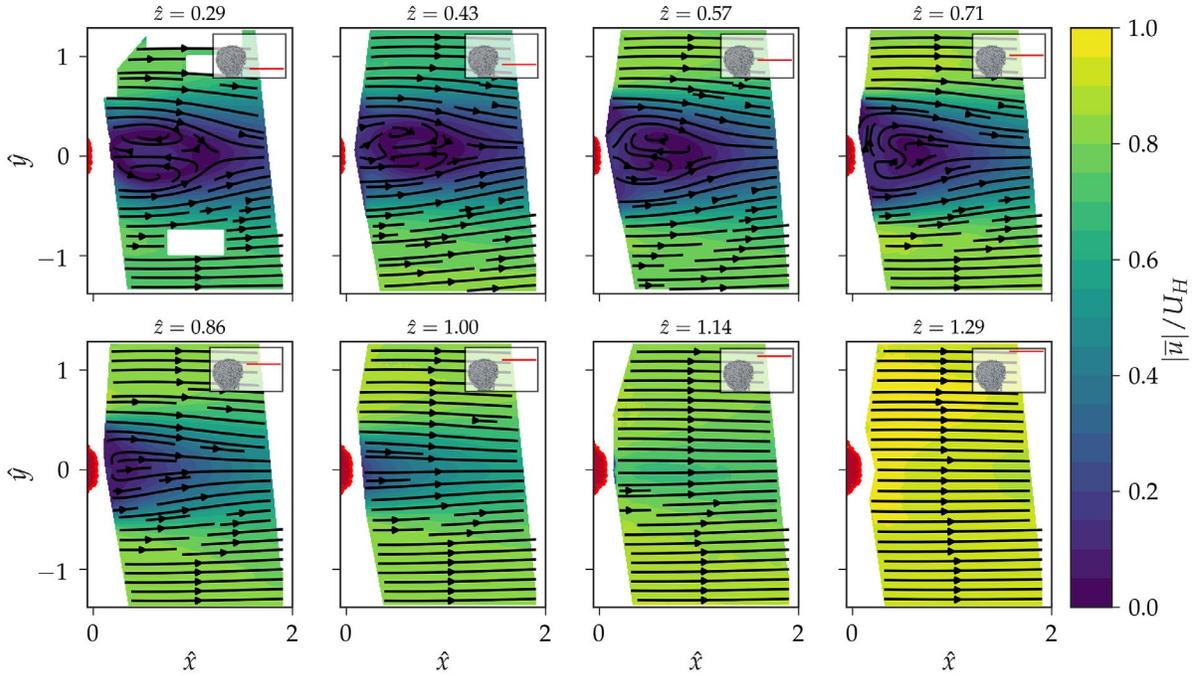

**Fig. 10** Normalized mean velocity magnitude $|\overline{\boldsymbol{u}}|/U_H$ at 8 horizontal planes, $\hat{z} = [0.29, 0.43, 0.57, 0.71, 0.86, 1.0, 1.14, 1.29]$ where the red shaded area indicates the location of the tree. The horizontal plane location is indicated by the red line in the in the sub-box.

### 3.2.2 Turbulent kinetic energy

An important impact of trees on the airflow is the modification of the turbulent kinetic energy (TKE). In literature, several studies emphasize the role of vegetation in turbulence enhancement for urban flows, which can drastically influence the pollution dispersion and thermal characteristics (Amorim et al., 2013; Gromke et al., 2008; Poggi et al., 2004). The internal structure of the plant is known to affect the budget of TKE directly. In addition to the well-known Richardson-Kolmogorov energy cascade from mean kinetic energy (MKE) to TKE (Pope, 2000), the plant is known to "*produce*" and "*dissipate*" the TKE due to the additional shortcut in energy transfer from MKE to TKE and the shortcut from TKE to turbulent dissipation rate (TDR) (Finnigan, 2000; Wilson and Shaw, 1977). This is due to the plant components interacting with the flow creating strong velocity gradients generating the TKE and the leaves act at turbulence suppressors, increasing the dissipation rate (Kenjereš and ter Kuile, 2013). The spatial variability in the turbulent kinetic energy (TKE) of the plant wake is investigated to understand the turbulence modification and link this effect to the plant porosity distribution. **Fig. 11** shows the normalized turbulent kinetic $k/U_H^2$ for the 8 horizontal planes. Trivially, the TKE is low above the tree ($\hat{z} > 1$) and directly behind the plant (near $\hat{x} = 0$, $\hat{y} = 0$ mm), where the air speeds are low. In contrast, the TKE is high at the shear zones between high-speed and low-speed zones. Comparing different planes, we see that the TKE profile is high near the ground ($\hat{z} = 0.29$) and near the plant canopy ($\hat{z} = 1$). This again is attributed to the high shear flows generated from the boundaries of the plant foliage that is present at these two vertical levels.

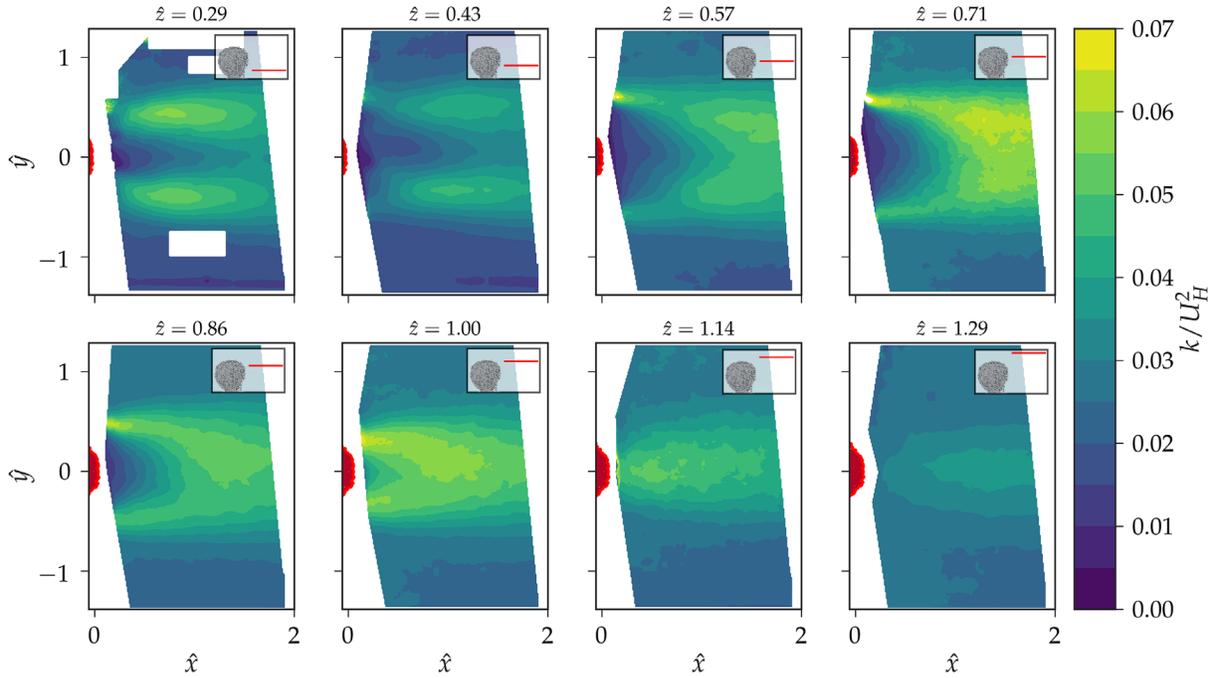

**Fig. 11** Normalized turbulent kinetic energy $k/U_H^2$ at 8 horizontal planes, $\hat{z} = [0.29, 0.43, 0.57, 0.71, 0.86, 1.0, 1.14, 1.29]$ where the red shaded area indicates the location of the tree. The horizontal plane location is indicated by the red line in the in the sub-box.

### 3.2.3 Linking flow field to plant porosity

To better study the influence of plant porosity on the wake flow, the centerline flow statistics and the plant porosity are compared. **Fig. 12** shows the vertical profiles of the center-line velocity at y = 0 for 7 streamwise positions, $\hat{x} = [0.4, 0.6, 0.8, 1.0, 1.2, 1.4, 1.6]$. The vertical profiles consist of streamwise velocity $\bar{u}/U_H$, the vertical velocity $\bar{w}/U_H$ and the turbulent kinetic energy $k/U_H^2$. Furthermore, to determine the relation between plant porosity distribution and wake flow field, the vertical streamwise-averaged porosity distribution $\langle\phi\rangle_x$ and aerodynamic porosity $\alpha$, obtained from **Fig. 8**b, is also plotted. We see that both the porosity distributions are only weakly linked to the streamwise velocity $\bar{u}/U_H$. This is observable at $\tilde{z} > 0.7$, as porosity approaches $\phi \to 1$, the streamwise velocity increases as well. In contrast, there is no apparent link between the vertical velocity $\bar{w}/U_H$ or turbulent kinetic energy $k/U_H^2$. Instead, we see that wake velocity deficit is simply governed by how the upstream wind profile in modified by the shear zones generated the outer geometry of the plant. This is because the ABL inflow condition has a more dominant influence on the wake profile and the resulting shear-layer and recirculation zone (as indicated by the negative streamwise velocity at $\hat{z} < 0.5$ and $\hat{x} < 0.8$). Therefore, in the present study, we do not observe a direct link between the aerodynamic porosity $\alpha$ nor the plant porosity $\phi$ with the wake flow statistics. Instead, it is dependent more on the total porosity of the plant and the geometry of the plant foliage.

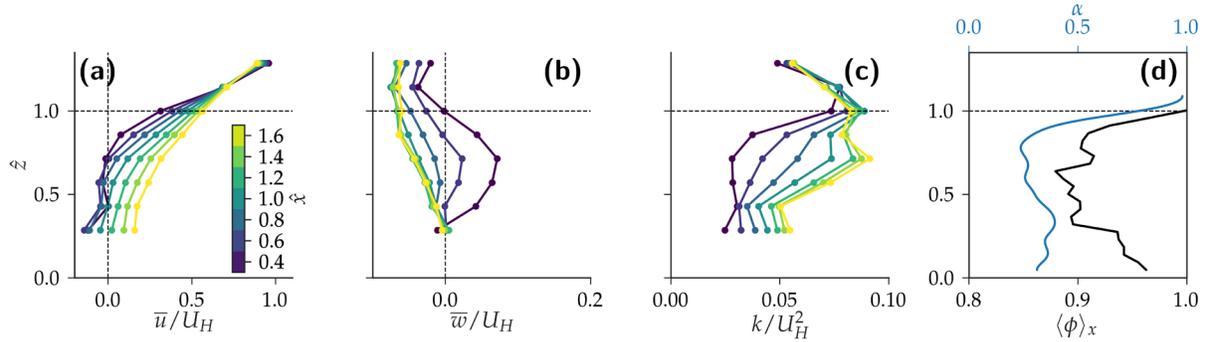

**Fig. 12** Mean normalized vertical profiles at 7 streamwise positions $\hat{x}$, at center-line of the plant $\hat{y} = 0$: (a) streamwise velocity $\bar{u}/U_H$ (b) vertical velocity $\bar{w}/U_H$, (c) turbulent kinetic energy $k/U_H^2$ and (d) streamwise-averaged porosity $\langle\phi\rangle_x$ and aerodynamic porosity $\alpha$.

## 3.3 Diurnal hygrothermal behavior of the plant

A thorough understanding of the plant morphology from **section 3.1** and the resulting flow characteristics in **section 3.2** enables us to link plant porosity to the wake flow characteristics. In this section, we investigate the link between plant morphology and the hygrothermal parameters such as air temperature and relative humidity. First, the diurnal response of the plant and its daytime and nighttime averages are investigated. Thereafter, the dynamic characteristics during the transition between day and night are investigated.

### 3.3.1 Influence of wind speed on the diurnal response

The diurnal behavior of the plant is investigated for two distinctly different boundary conditions: *no-wind* condition and *wind* condition with $U_{ref} = 1$ m s$^{-1}$. **Fig. 13** shows the diurnal variation of the water mass loss $m$ (g) throughout the day and night and the resulting the transpiration rate $TR$ (g h$^{-1}$), defined as:

$$TR = \frac{dm}{dt} \qquad (5)$$

measuring the hourly change in mass due to transpiration. We observe that, during the night regardless of the wind condition, a constant transpiration rate of 2.5 g h$^{-1}$ exists. This transpiration rate, in the absence of solar radiation, is therefore associated to the water loss due to dark respiration (Farquhar et al., 1980; Lambers et al., 2008; Launiainen et al., 2015). At dawn, a strong increase in transpiration rate is observable. Furthermore, at this time, the wind speed plays an important role: with the wind, a peak transpiration rate of 15 g h$^{-1}$ is observed whereas, without wind, the transpiration rate peaks only at 10 g h$^{-1}$. As the day progresses, the stomatal regulation is seen to compensate for the influence of wind, resulting in similar transpiration rate with an average transpiration rate of 9 g h$^{-1}$. The decay in the daytime transpiration after dawn has also been observed previously (Javaux et al., 2013; Tuzet et al., 2003). The phenomenon is attributed to reducing rhizosphere soil moisture content. The reduced soil moisture eventually reduces the stomatal conductance, resulting in the decayed transpiration rate that we observe. However, at night, the soil moisture equilibrates as the root water uptake is drastically diminished to only 2.5 g h$^{-1}$.

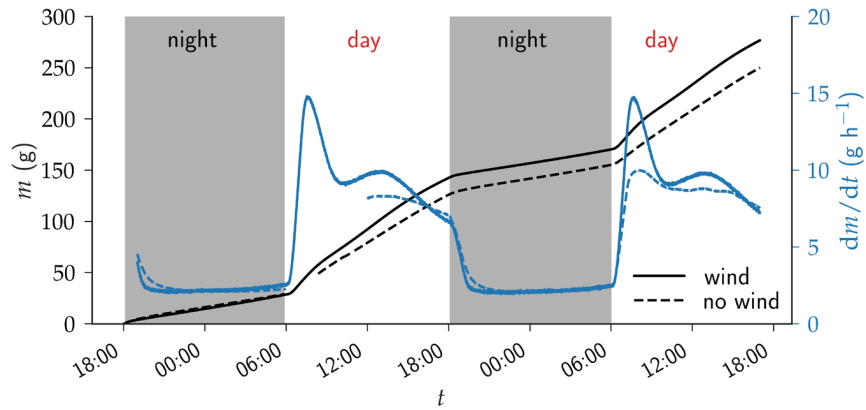

**Fig. 13** Diurnal variation of mass loss $m$ (g) and the resulting transpiration rate $TR = \mathrm{d}m/\mathrm{d}t$ (g h$^{-1}$) for two wind conditions: *no-wind* and *with* wind ($U_{ref} = 1$ m s$^{-1}$). Note: A part of *no-wind* data is missing (between 06:00 to 12:00) due to a fault in data acquisition.

The influence of the varying transpiration rate is further investigated by studying the hygrothermal microclimate variables, air temperature $T$ (°C) and relative humidity $RH$ (%) inside the plant foliage. The setup of the hygrothermal sensors was detailed in **section 2.2.2**. **Fig. 14** shows the diurnal variation of the air temperature and relative humidity inside the tree at various heights, with $T_1$ at the bottom of foliage and $T_6$ at the plant canopy ( $\hat{z} = 1$). The configuration is such that probes 1 to 5 are directly inside the plant foliage with 30 mm offset. Furthermore, the ambient condition ("*air*") and the ground condition below the plant ("*ground*") is compared to study the influence of plant shading. **Fig. 14** shows that, in the *no-wind* condition, there is a quantifiable drop over height in the air temperature and a substantial increase in the relative humidity, with peak $RH = 75\%$ during day time. During the night, the vertical variability in the environmental conditions within the foliage is smaller but still noticeable. All the sensors inside the plant foliage show lower temperatures during day and night. However, the sensor at plant canopy height ($\hat{z} = 1$), shows that the air temperature is noticeably higher than the ambient condition. This indicates that the plant canopy region is strongly influenced by the absorption of solar radiation, leading to higher leaf temperature and thereby heating up the air. Similar observation of such positive leaf-to-air temperature due to high solar radiation absorption have also been observed numerically (Manickathan et al., 2018b). However, in the shadow of the plant, we see that air temperature is lower than sunlight region. We must note that the soil is sealed so that water loss is simply due to transpiration of the leaves. Therefore, the cooling by the ground is attributed to the plant shading only resulting in less energy storage and leading to lower air temperatures.

The impact of transpiration, i.e., local cool "*oasis*", diminishes strongly when the wind is present. The increased ventilation of the plant foliage due to the wind is seen to reduce the cool oasis formed by the transpirative cooling indicated by the higher air temperature compared to the no wind conditions especially in the lower zone of the tree. This ventilation by wind also leads to lower relative humidity values in the tree. **Fig. 13** shows that the influence of wind on transpiration rate is, however, more complex. When the relative humidity inside the plant is lower, the vapor pressure deficit (VPD) will also be higher (Manickathan et al., 2018b). Thus, at dawn, due to the higher atmospheric evaporative demand (AED), the transpiration rate is substantially higher than in the *no-wind* case. This observation is in line with the theory that transpiration scales with atmospheric evaporative demand (McVicar et al., 2012). At dawn, when the stomatal resistance is low (or stomatal conductance is high), the Penman-Monteith equation predicts an increasing transpiration rate with increasing wind speed (Idso, 1977; Nobel, 2009). However, as the day progresses, the transpiration rate is seen to equilibrate down to the *no-wind* condition level. This is the result of the influence of wind speed on the leaf water use efficiency (WUE, i.e., carbon uptake rate for a given transpiration rate), resulting in a decreasing transpiration rate (Dixon and Grace, 1983; Nobel, 2009; Schymanski and Or, 2016). As the day progresses, due to the high sensible heat flux providing efficient convective cooling to the leaves, it improves the plant's ability to conserve water, reducing the stomatal conductance (Idso, 1977). Thus, even though there is higher wind speed and an equivalently higher AED, the transpiration rate is seen to be similar to that of the *no-wind* condition.

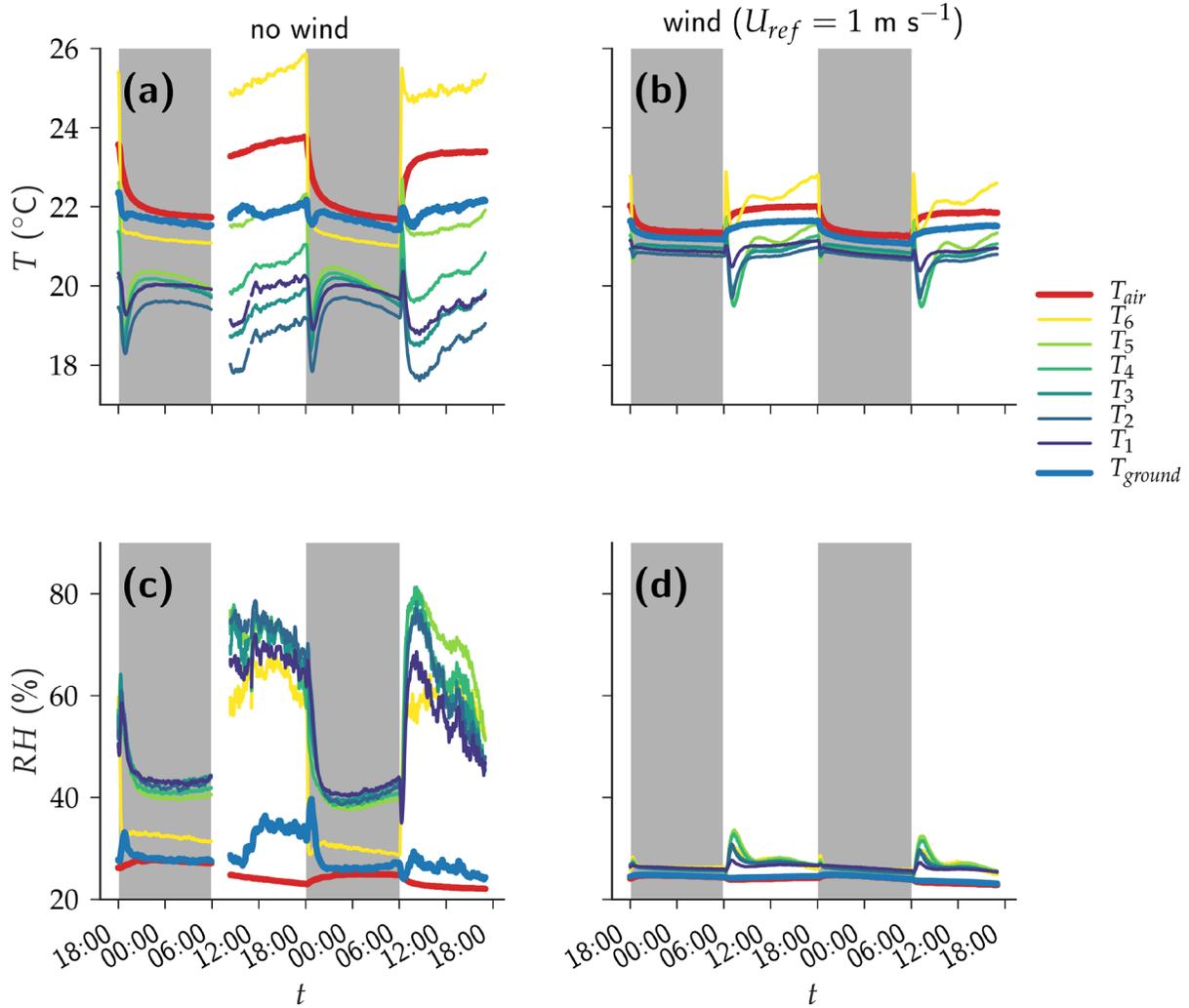

**Fig. 14** Diurnal variation of air temperature *T* (°C) and relative humidity *RH* (%) inside the tree at varying heights for two wind conditions: (a)(c) *no-wind* and (b)(d) *wind* condition ($U_{ref} = 1$ m s$^{-1}$).

### 3.3.2 Day versus night comparison of plant response

The average daily response of the plant and its influence on the climate are studied to understand the cooling provided by the plant. **Fig. 15**(a-d) shows the median air temperature and relative humidity at various heights. **Fig. 15**(a-b) shows *no-wind* case and **Fig. 15**(c-d) shows the wind case. Furthermore, the figure differentiates the daytime (i.e., 08:00 to 16:00) and nighttime (i.e., 20:00 to 04:00) median air temperature (i.e., a temporal-averaging window size of 8 hours). The daytime and nighttime period were shortened to focus on the period that is nearly steady-state. The figure shows that the air temperature change is largest during the day. However, more interestingly, the figure shows that, in the presence of wind, the air temperature inside the tree is more homogenized, whereas, without wind, a large vertical variability is observed. We see that the vertical profile of temperature and relative humidity is correlated with the leaf area density $\langle a \rangle_{xy}$ and solar radiative heat flux $\langle q_{r,sw} \rangle_{xy}$, especially during *no-wind* condition, shown in **Fig. 15**e. The most significant cooling, i.e., providing the largest drop in air temperature, is present for the *no-wind* condition during the day for heights between $0.29 \leq \hat{z} \leq 0.57$. Moreover, at these regions, we also observe a large increase in the relative humidity. We see that relative humidity is directly correlated with the vertical profile of leaf area density. At regions of high LAD, there is a higher transpiration rate resulting in a higher transpirative cooling. Moreover, the enhanced effect of transpirative cooling and the generation of the local cool zone is seen to be only prevalent when the wind is low.

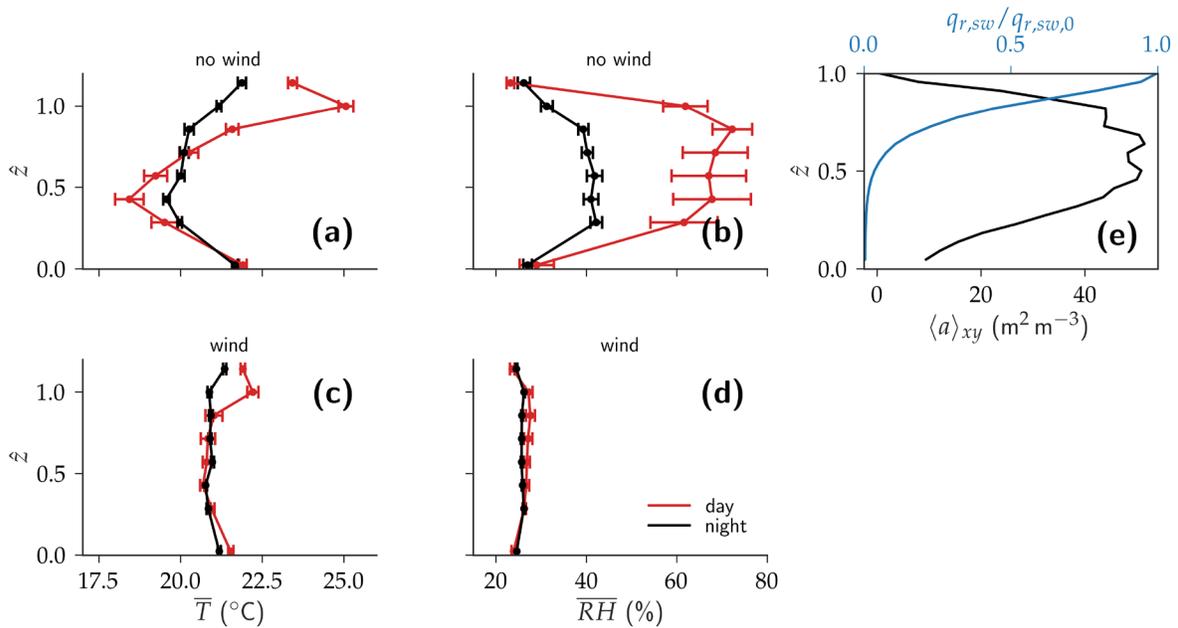

**Fig. 15** Mean vertical distribution of day (08:00 – 16:00) (red) and night (20:00 – 04:00) (black) (a)(c) air temperature $T$ (°C) and (b)(d) relative humidity $RH$ (%) inside the tree for two wind conditions: (a)(b) *no-wind* and (c)(d) *wind* condition ($U_{ref} = 1$ m s$^{-1}$); and (e) horizontal-averaged vertical distribution of leaf area density $\langle a \rangle_{xy}$ (m$^2$ m$^{-3}$) and normalized short-wave radiation intensity $\langle q_{r,sw}/q_{r,sw,0} \rangle_{xy}$.

High air temperature is observed at the plant canopy region and is the result of high solar radiation absorption at this layer, as indicated by the solar radiative heat flux profile (**Fig. 15**e) As most of the solar radiation is absorbed in the plant canopy (i.e., top 20% of the foliage), it can result in a large increase in leaf temperature (i.e., $T = 25$ °C) and thereby a strong increase in the air temperature (Manickathan et al., 2018b). Such gradient in radiation absorption results in the observed air temperature spike at plant canopy as indicated by probe 6 ($\hat{z} = 1$). In contrast, for probes 1 to 5 ($\hat{z} = [0.29, 0.44, 0.57, 0.71, 0.86]$) that are within the foliage, which are at regions of low solar radiation intensity, significantly lower air temperature is observed.

To investigate the variation in leaf temperature and its link to transpiration and solar radiation, the leaf temperature at the exterior of the plant is measured through infrared thermography. **Fig. 16** shows the diurnal variation of leaf surface temperature for *wind* and *no-wind* condition. The thermography shows that, during the *wind* condition, the spatial variability in leaf temperature is minimal, especially during the night. Therefore, the convective cooling of the leaves homogenizes the vertical variability in the leaf temperature. This is especially evident when observing the *no-wind* condition. During the *no-wind* condition, the vertical variation in leaf temperature shows a high plant-canopy leaf temperature and a cool in-foliage leaf temperature, especially at the lower regions. This observation reflects the hygrothermal measurements inside the foliage (**Fig. 15**(a-d)), and the vertical distribution of the solar radiative heat flux (**Fig. 15**e). Where solar radiation intensity is high, there is a large increase in the leaf temperature. So, we see that solar radiation is largest contributor to the leaf temperature rise. At night (**Fig. 16**a and **Fig. 16**b), the spatial variability is negligible. Although, the wind is seen to reduce the leaf temperature due to increased convective heat transfer, as indicated by **Fig. 16**c and **Fig. 16**f.

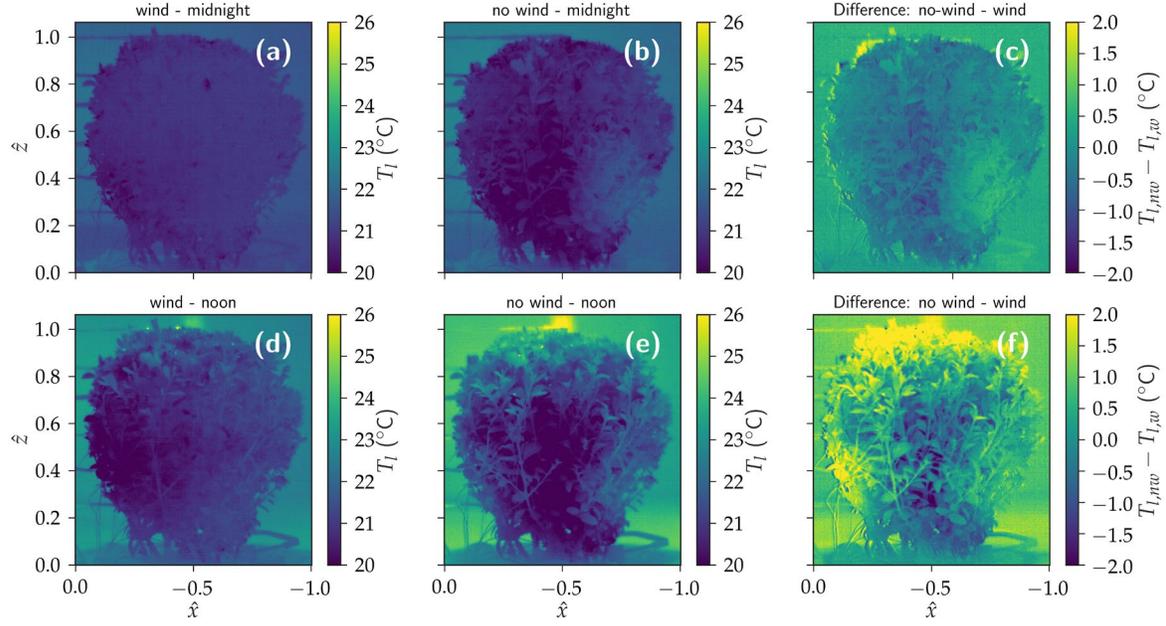

**Fig. 16** Diurnal variation of leaf surface temperature $T_l$ (°C) at (a)(b)(c) nighttime (midnight) (d)(e)(f) and midday (noon). The difference between *no-wind* and *wind* condition is compared for (c) night and (f) day.

During the transition between day and night, a faster and a stronger variability is observed, influenced by the abrupt change in the environmental condition. **Fig. 17** shows the spatiotemporal variability of streamwise-averaged leaf temperature $\langle T_l \rangle_x$ for *no-wind* and *wind* conditions. The figure shows the vertical variability in leaf temperature over the progression of the day. The streamwise averaging is only performed on the pixels with foliage. We see that, as observed with the internal hygrothermal microclimate of the plant (**Fig. 15**), the variability is only prevalent during the daytime. Furthermore, the *no-wind* condition shows stronger temporal and spatial change than the *wind* condition. We see that, during the *wind* condition, the enhanced convective cooling provided by the wind reduces the high leaf temperature at the plant-canopy where the solar radiation penetrates. Furthermore, we observe that the strongest cooling is observed in the middle regions of the plant foliage around dawn with $T_l \approx 19$ °C. This indicates that leaf transpiration is also present at parts of foliage where solar radiation is lower. As the day progress, however, we see a diminishing cooling effect from the plant, indicated by the rising foliage temperature. To better understand this temporal variation between the two wind conditions, we investigate the net spatially-averaged leaf temperature history. It must be noted that the leaf temperature that is measured using the infrared imaging is simply the outer plant foliage temperature. The leaf temperature inside the plant foliage is known to be different and closer to the air temperature (Manickathan et al., 2018b). The air temperature inside the plant foliage is seen to be lower and more uniform as indicated from the hygrothermal measurement, as shown in **Fig. 15**.

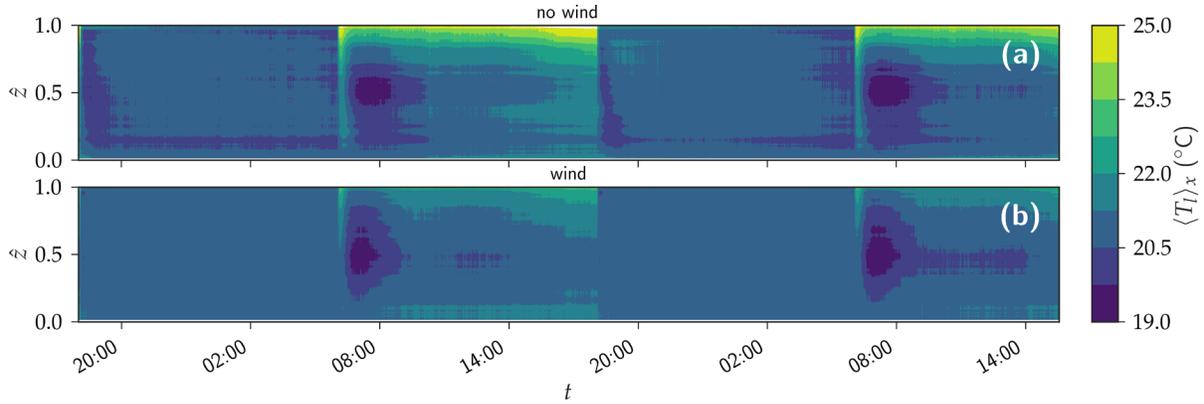

**Fig. 17** Streamwise-averaged spatiotemporal variability of leaf temperature $\langle T_l \rangle_x$ (°C) comparing (a) *no-wind* and (b) *wind* condition. The averaging operator is performed with the pixels belonging to the foliage.

**Fig. 18** shows the diurnal evolution of the net spatial-averaged leaf temperature $\langle T_l \rangle_{xz}$ (i.e., average pixel-value of all pixels belonging to plant foliage shown in **Fig. 16**) comparing *wind* and *no-wind* conditions. We notice that the *wind* condition and *no-wind* condition responses are very similar with the *no-wind* condition being more amplified. Furthermore, we see that the day is composed of four unique stages: "*no-cooling*" stage, "*high-cooling*" stage, "*equilibrium*" stage, and finally "*decaying-cooling*" stage. The *no-cooling* period is present at the initial stages of dawn where the leaves absorb solar radiation and the stomatal response has not been prevalent to provide an adequate transpiration rate. The delayed response of the plant results in the high overshoot of leaf temperature, which is compensated and corrected by the plant thereafter. However, we see that, during the *high-cooling* stage, the transpiration rate spikes, as evident from plant transpiration rate measurement, **Fig. 13**, resulting in the drastic cooling measured. By midday, the stomatal response equilibrates the necessary transpiration rate and we observe a quasi-steady leaf temperature and transpiration rate (**Fig. 13**) and in-foliage air temperature (**Fig. 14**). Furthermore, the equilibrium leaf temperature of plant foliage during the *no-wind* condition is higher than for the *wind* condition. This is the result of the overall higher plant canopy temperature due to the reduced convective heat transfer. As the day progresses, the leaf temperature transitions to the fourth stage (i.e., *decaying-cooling* stage) with a slow increase in the leaf temperature. The observation correlates with the measured plant transpiration rate, **Fig. 13**, where an equally decaying transpiration rate is observable.

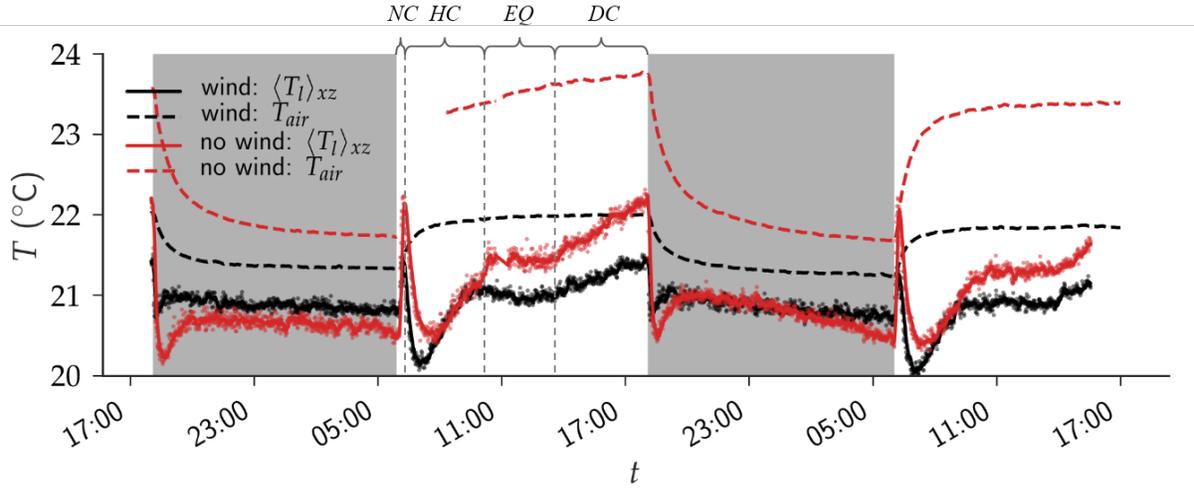

**Fig. 18** Diurnal variation of spatially-averaged leaf temperature $\langle T_l \rangle$ (°C) for *wind* and *no-wind* conditions. The dashed-line indicates the ambient air temperature $T_{air}$ (as shown in **Fig. 14**) measured at $\hat{z} = 1$. The day is comprised of four stages: *no-cooling (NC), high-cooling (HC), equilibrium (EQ),* and *decaying-cooling (DC)* stages.

**Fig. 19** shows a diurnal cycle of the net leaf temperature versus plant transpiration rate hysteresis showing hysteresis: i.e., a delayed response resulting in a temporal cyclic pattern. **Fig. 19**a and **Fig. 19**c shows the hysteresis of plant transpiration rate with respect to plant canopy region leaf temperature for *no-wind* and *wind* condition, respectively. Whereas, **Fig. 19**b and **Fig. 19**d shows the hysteresis with respect to ground-region leaf temperature. We see that the hysteresis magnitude at the bottom is less amplified in contrast to the plant canopy leaf temperature. Furthermore, we observe that leaf temperature rises rapidity at the start of the day due to the absorption of solar radiation. After a delay, the transpiration helps cool the leaves and reaches a quasi-equilibrium between transpiration rate and leaf temperature. In the absence of solar radiation (i.e., after dusk), the leaf temperature quickly decreases. However, we observe that the transpiration rate is not reduced at the same rate. Only after a delayed time-period, does the transpiration rate reduce, resulting in a slight increase in leaf temperature and reaching the night-time equilibrium state. This diurnal circular cyclic pattern of leaf temperature and transpiration rate, instead of simple linear variation, is observable due to delay in the stomatal response. The hysteresis occurs because the stomatal conductance is not just dependent on the atmospheric evaporative demand (AED) such as light, temperature, and humidity, but also depends on the water transport within the xylem. The root water uptake and the resulting water transport through the plant is known to have a delayed response due to the capacitance on the sub-elements in the plant leaf-xylem-root system. The hysteresis due to the mismatch in transpiration demand and the root water uptake has been well documented in the past (Dauzat et al., 2001; Williams and Rastetter, 1996). An essential consequence of this dynamic plant response is that it is impossible to correctly parameterize vegetation in urban climate models without explicit modeling the water transport within the plant. Therefore, a higher-fidelity model, modeling the dynamic water transport in the plant system is required to capture such hysteresis patterns (Huang et al., 2017; Manzoni et al., 2011). Furthermore, a simpler model might not be able to predict such rapid changes in leaf temperature when the environmental conditions abruptly change. For example, solar radiation might rapidly change due to changes in cloud cover or the changes in solar shading from nearby buildings, and rapid changes in AED due to changes in wind speed (such as gusts).

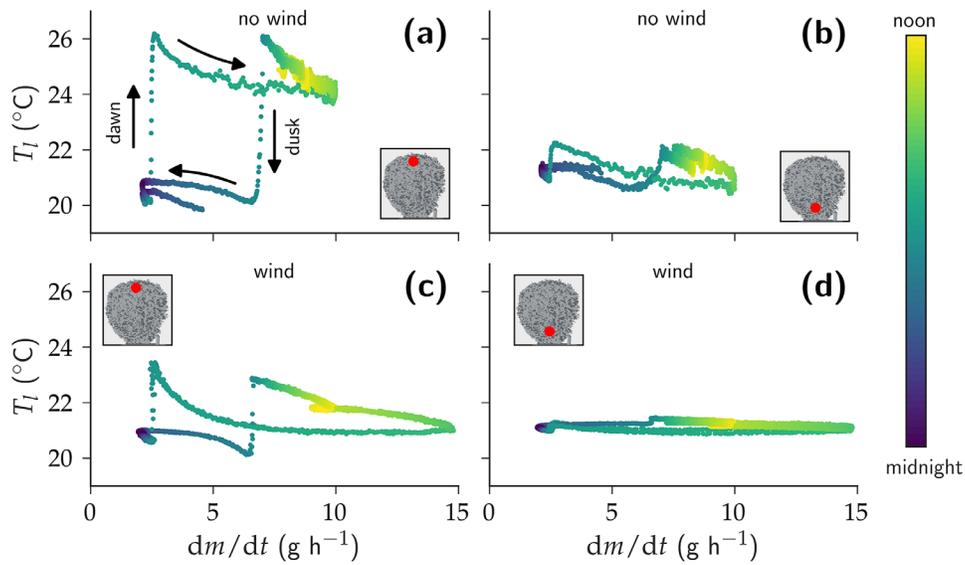

**Fig. 19** Diurnal variation of $TR = dm/dt$ (g h$^{-1}$) and leaf temperature for two wind conditions: (a)(b) *no-wind* and (c)(d) *wind* condition ($U_{ref} = 1$ m s$^{-1}$). The leaf temperatures are obtained from (a)(c) plant canopy region and (b)(d) near the ground region of the foliage. The red point on the plant indicates the location where the leaf temperature sensor was located.

# 3. Conclusion

The goal of the present study was to unveil the diurnal changes in plant microclimate using multiple non-intrusive imaging techniques such as stereoscopic particle image velocimetry (SPIV) for the flow field, infrared thermography for the leaf temperature and X-ray tomography for the plant microstructure. The present study aimed at answering the following questions: What are the spatial and temporal variability of the plant performance due to environmental conditions such as wind speed and solar radiation? What is the diurnal response of the plant?

The high-resolution measurement of the plant porosity through X-ray tomography enabled us to find that that the aerodynamic and optical porosities typically used in the wind tunnel studies do not reflect the true porosity distribution of the plant. Moreover, the paper presents a novel approach of determining the leaf area density directly from X-ray tomography. The advantage of this non-intrusive approach of determining the plant microstructure enabled us to fit series of additional experiments, enabling us to directly quantify the impact of plant foliage morphology on the wake flow characteristics, the hygrothermal conditions such as air temperature and relative humidity inside the plant foliage, the solar radiation penetration inside the foliage and, finally, its impact on the spatial distribution of the leaf temperature. The SPIV measurement of the wake flow field helped us to find that at regions where porosity $\phi \to 1$, a strong bleed flow is observable, indicated by the mean velocity component. However, in contrast, there was no apparent link between the plant porosity distribution and the turbulent kinetic energy (TKE). The TKE intensity was seen to be governed by the net plant porosity and the outer geometry of the plant foliage that generates the shear-layer and how it interacts with the upstream boundary layer profile.

The hygrothermal measurement at multiple vertical locations within the plant foliage enabled us to find that the local cool zone generated by transpirative cooling quickly diminished when the wind is present. The diurnal measurement of the transpiration rate showed that the water use efficiency (WUE) changes during the day indicated by the decaying transpiration rate. Furthermore, the high-resolution infrared thermography measuring the spatial and temporal changes in the leaf temperature revealed further dynamic variability during the day. A comparison of the diurnal variation in the leaf temperature and the net plant transpiration rate enabled us to quantify the diurnal hysteresis resulting from the stomatal response lag. The plant day is seen to comprise of four unique stages *no-cooling* (i.e., the stage when stomata has not responded to the increase in solar radiation), *high-cooling* (i.e., when stomatal response tries to compensate for increased leaf temperature), *equilibrium* (i.e., when stomatal response and leaf temperature equilibrates) and *decaying-cooling* stage (i.e., when the transpiration rate starts to weaken). Such plant responses are difficult to parameterize and simplify for urban climate-vegetation models without explicitly modeling the water transport within the plant. The challenge of such complex dynamics is that simplified models might not be able to predict rapid changes in leaf temperature due to the sudden change in atmospheric evaporative demand (AED) resulting from a sudden change in solar radiation (e.g. due to a sudden change in cloud cover) or sudden change in wind speed (e.g. due to gust). Therefore, higher-fidelity models of plant responses should take into account such dynamics that arise from water availability and the stomatal response delay, to accurately assess the transpirative cooling potential of vegetation. A further contribution of the present paper is to provide high-resolution multivariate measurement dataset for development and validation towards such advanced numerical methods.

## Acknowledgments

L.M. gratefully acknowledges the support of Kevin Prawiranto for preliminary X-ray tomography, Robert Fischer for fruitful discussions on image segmentation, Roger Vonbank and Beat Margelisch with the wind tunnel setup, Jiggar Shah with SPIV calibration and PIV setup, Christina Tsalicoglou with the setup of boundary layer generators and Stephan Carl for the full-setup design and implementation. L.M. would also like to gratefully acknowledge the contribution of Dr. Richter from UZH and Rolf Kaufmann from Empa, for helping and providing the opportunity for performing the X-ray CT scan. The funding for L.M's Ph.D. is provided by the Chair of Building Physics at ETHZ.